\documentclass[aps,preprint,nofootinbib,showpacs]{revtex4}
\usepackage{graphicx,color,amsmath, epsfig}
\newcommand{\be}{\begin{equation}}
\newcommand{\ee}{\end{equation}}
\newcommand{\ba}{\begin{eqnarray}}
\newcommand{\beq}{\begin{equation}}
\newcommand{\eeq}{\end{equation}}
\newcommand{\ea}{\end{eqnarray}}
\newcommand{\gev}{\, {\rm GeV}}
\def\bea{\begin{eqnarray}}
\def\eea{\end{eqnarray}}

\def\err#1#2{\lower2pt\hbox{ $\stackrel{\scriptstyle +#1}{\scriptstyle -#2}$}}
\def\ga{\mathrel{\raise.3ex\hbox{$>$\kern-.75em\lower1ex\hbox{$\sim$}}}}
\def\la{\mathrel{\raise.3ex\hbox{$<$\kern-.75em\lower1ex\hbox{$\sim$}}}}

\begin{document}

\preprint{%
\vbox{%
\hbox{TUM-HEP-683/08}
}}
\title{\boldmath Measuring $B^\pm\to \tau^\pm\nu$ and $B^\pm_c\to \tau^\pm\nu$ 
at the $Z$ peak \unboldmath}

\author{A.G. Akeroyd$^{1,2}$, Chuan Hung Chen $^{1,2}$, S. Recksiegel$^{3}$}
\affiliation{$^1$ Department of Physics, National Cheng Kung University, 
Tainan 701, Taiwan} 
\affiliation{$^2$ National Center for Theoretical Sciences, Taiwan}
\affiliation{$^3$ Physikdepartment T31, Technische Universit\"at M\"unchen,
D-85747 Garching, Germany}

\date{\today}
\begin{abstract}
The measurement of $B^\pm\to \tau^\pm\nu_\tau$ at the $B$ factories provides
important constraints on the parameter $\tan\beta/m_{H^\pm}$ in the
context of models with two Higgs doublets.
Limits on this decay from 
$e^+e^-$ collisions at the $Z$ peak were sensitive to the sum of
$B^\pm\to \tau^\pm\nu_\tau$ and   $B^\pm_c\to \tau^\pm\nu_\tau$.
Due to the possibly sizeable contribution from $B^\pm_c\to \tau^\pm\nu_\tau$
we suggest that a signal for this combination might be observed 
if the LEP L3 Collaboration used their total data of 
$\sim 3.6\times 10^6$ hadronic decays of the $Z$ boson.
Moreover, we point out that a future Linear Collider operating at the $Z$ peak
(Giga $Z$ option) could constrain $\tan\beta/m_{H^\pm}$ 
from the sum of these processes with a precision comparable 
to that anticipated at proposed high luminosity $B$ factories from
$B^\pm\to \tau^\pm\nu_\tau$ alone.
\end{abstract}
\maketitle

\section{Introduction}
In April 2006 the BELLE collaboration announced the first observation of the
purely leptonic decay $B^\pm\to \tau^\pm \nu_\tau$ \cite{Ikado:2006un}
utilizing an integrated luminosity of $414$ fb$^{-1}$. The measured
branching ratio (BR) is in agreement with the Standard Model (SM) rate
within theoretical and experimental errors:
\begin{equation}
{\rm BR}(B^\pm\to \tau^\pm\nu_\tau)=
\left(1.79\,^{+0.56}_{-0.49}(stat)\,^{+0.46}_{-0.51}(syst)\right)
\times 10^{-4}
\label{exp_btaunu}
\end{equation}
Subsequently, the BABAR collaboration reported a measurement with an 
integrated luminosity of  $346$ fb$^{-1}$
which is an average
of separate analyses with semi-leptonic  \cite{:2007bx} 
and hadronic \cite{:2007xj} tags:
\begin{equation}
{\rm BR}(B^\pm\to \tau^\pm\nu_\tau)= \left(1.2 \pm 0.4\pm 0.3 \pm 0.2 \right)\times 10^{-4}
\end{equation}
The average of the BELLE and BABAR measurements is \cite{HFAG}:
\begin{equation}
{\rm BR}(B^\pm\to \tau^\pm\nu_\tau)= \left(1.41\,^{+0.43}_{-0.42} 
\right)\times 10^{-4}
\end{equation}
Significantly improved precision for BR$(B^\pm\to \tau^\pm\nu_\tau)$
would require a high luminosity ${\cal L}\ge 10^{35}cm^{-2}s^{-1}$ 
$B$ factory
\cite{Yamauchi:2002ru,Bigi:2004kn,Hashimoto:2004sm,Browder:2004wu,
Hewett:2004tv,Bona:2007qt,Browder:2008em}.
In the context of the SM the decay $B^\pm\to \tau^\pm\nu_\tau$ provides 
a direct measurement of the combination $f_BV_{ub}$, where
$f_B$ is the decay constant which can only be calculated by
non-perturbative techniques such as lattice QCD. 
Charged Higgs bosons ($H^\pm$) present in the 
Two Higgs Doublet Model (2HDM) and the Minimal Supersymmetric SM (MSSM)
would also mediate $B^\pm\to \tau^\pm\nu_\tau$ \cite{Albright:1979yc}
with the New Physics contribution being sizeably enhanced if $\tan\beta$ 
(the ratio of vacuum expectation values of the two Higgs doublets) 
is large \cite{Hou:1992sy}.
The above measurements of $B^\pm\to \tau^\pm\nu_\tau$ now
provide a very important constraint on the parameter $\tan\beta/m_{H^\pm}$
in the context of the 2HDM and the MSSM.
Hence this decay is of much interest in both the SM and
models beyond the SM and improved precision in the above
measurements is certainly desirable. 

Prior to the era of the $B$ factories
three LEP collaborations searched for $B^\pm\to \tau^\pm\nu_\tau$ and obtained upper bounds
within an order of magnitude of the SM prediction \cite{Acciarri:1996bv,Abreu:1999xe,Barate:2000rc}.
Such limits were actually
sensitive to the sum of $B^\pm\to \tau^\pm\nu_\tau$ and $B^\pm_c\to \tau^\pm\nu_\tau$
\cite{Mangano:1997md} since the centre-of-mass energy 
($\sqrt s=91$ GeV) was above the $B_c^\pm$ production 
threshold (unlike the $B$ factories). The strongest limits were set by the L3 collaboration which obtained
BR($B^\pm\to \tau^\pm\nu_\tau)< 5.7\times 10^{-4}$ \cite{Acciarri:1996bv}. Since
BR($B^\pm\to \tau^\pm\nu_\tau$) has now been measured at the $B$ factories, the L3 limit 
can now be used to provide a limit on 
the product of the transition probability $f(b\to B_c)$ and
BR($B^\pm_c\to \tau^\pm\nu_\tau$). A quantitative study of the
magnitude of the contribution of 
$B^\pm_c\to \tau^\pm\nu_\tau$ to the LEP limits
was performed in \cite{Mangano:1997md}. We update this analysis using the
significant improvements in the measurements of the CKM matrix and calculations of $f_B$.
Moreover, the measurements of the $B_c^\pm$ production cross-section 
at the Fermilab Tevatron \cite{Abe:1998fb,Abulencia:2005usa,Abulencia:2006zu,
Aaltonen:2007gv}
provide the first measurements of the transition probability for $b\to B_c$
and suggest much larger values than the theoretical estimations
used in the numerical analysis of \cite{Mangano:1997md}.
The L3 limit on BR($B^\pm_c\to \tau^\pm\nu_\tau$)
was obtained with $\sim 1.5\times 10^6$ hadronic $Z$ boson decays,
 which is slightly less than half the full L3 data taken at the $Z$ peak. 
We suggest that a search for $B^\pm/B^\pm_c\to \tau^\pm\nu_\tau$
using the full L3 data sample ($\sim 3.6\times 10^6$ hadronic 
$Z$ decays \cite{:2005ema}) would not only
strengthen the limit on the product of 
$f(b\to B_c)$ and BR($B^\pm_c\to \tau^\pm\nu_\tau$)
but also offer the possibility of a signal, which would be an additional 
observation of $B^\pm\to \tau^\pm\nu_\tau$ and
the first observation of $B^\pm_c\to \tau^\pm\nu_\tau$.
It is also pointed out that a future $e^+e^-$ Linear Collider operating
at the $Z$ peak (the Giga $Z$ option \cite{Hawkings:1999ac,
AguilarSaavedra:2001rg,Ali:2000cq,Djouadi:2007ik}) 
could offer similar sensitivity to the parameter $\tan\beta/m_{H^\pm}$
from these leptonic decays
as the proposed high luminosity $B$ factories.
This article is structured as follows: in section II we present
basic formulae for the decay rates for
$B^\pm/B_c^\pm\to \tau^\pm\nu_\tau$ and discuss the $H^\pm$
contribution; we study the admixture of $B^\pm_c\to \tau^\pm\nu$
and $B^\pm\to \tau^\pm\nu$ at the $Z$ peak in section III 
and give our conclusions in section IV.
\boldmath
\section{The decays $B^\pm\to \tau^\pm\nu$ and  $B^\pm_c\to \tau^\pm\nu$}
\unboldmath

In the SM, the purely leptonic decays ($\ell^\pm\nu_\ell$) 
of $B^\pm$ and $B_c^\pm$ proceed via 
annihilation to a $W$ boson in the $s$-channel. The decay rate is given by
(where $q=u$ or $c$):
\begin{equation}
\Gamma(B^+_q\to \ell^+\nu_\ell)=\frac{G_F^2 m_{B_q} m_\ell^2 f_{B_q}^2}{8\pi}
|V_{qb}|^2 \left(1-\frac{m_\ell^2}{m^2_{B_q}}\right)^2
\end{equation}
Due to helicity suppression, the rate is proportional to $m^2_\ell$
and one expects:
\begin{equation}
BR(B^+_q\to \tau^+\nu_{\tau}):BR(B^+_q\to \mu^+\nu_{\mu})
:BR(B^+_q\to e^+\nu_{e})=m^2_{\tau}:m^2_{\mu}:m^2_e
\end{equation}
These decays are relatively much more important for $B^\pm_c$
than $B^\pm_u$ due to the enhancement factor $|V_{cb}/V_{ub}|^2
(f_{B_c}/ f_{B_u})^2$. Using the input parameters given in
table \ref{parameters}, we obtain the SM predictions listed
in table \ref{SMBRs}.

\begin{table}
\center{\begin{tabular}{|l|l|}
\hline
 $G_F=1.16639\cdot 10^{-5} \gev^{-2}$ & $m_e=0.511\,{\rm MeV}$ \\
 $m_\mu=0.10566\gev$ & $m_\tau=1.777\gev$ \\\hline
 $|V_{ub}|=0.00386(28)$ &   $|V_{cb}|= 0.0416(9)$ \\\hline 
 $m_{B_u}= 5.279\gev$ & $\tau_{B_u}= 1.638\cdot 10^{-12}s$ \\
 $m_{B_c}= 6.271\gev$ & $\tau_{B_c}= 0.463(177)\cdot 10^{-12}s$ \\ 
 $f_{B_u}= 0.216(22)\gev$ & $f_{B_c}= 0.450 \gev $\\
\hline
\end{tabular} }
\caption{Input parameters used in this paper, unless indicated otherwise
in the text.}
\label{parameters}
\end{table}

\begin{table}
\center{\begin{tabular}{|l|ccc|}
\hline
& $BR(B^+_q\to \tau^+\nu_{\tau})$ & $BR(B^+_q\to \mu^+\nu_{\mu})$
&  $BR(B^+_q\to e^+\nu_{e})$\\\hline
$B_u$ & $1.2 \cdot 10^{-4}$ & $5.5 \cdot 10^{-7}$ & $1.3 \cdot 10^{-11}$ \\
$B_c$ & $0.022$             & $9.3 \cdot 10^{-5}$ & $2.2 \cdot 10^{-9}$ \\
\hline
\end{tabular} }
\caption{Standard Model predictions for the branching ratios (central values).}
\label{SMBRs}
\end{table}

\begin{figure}
\center{\epsfig{file=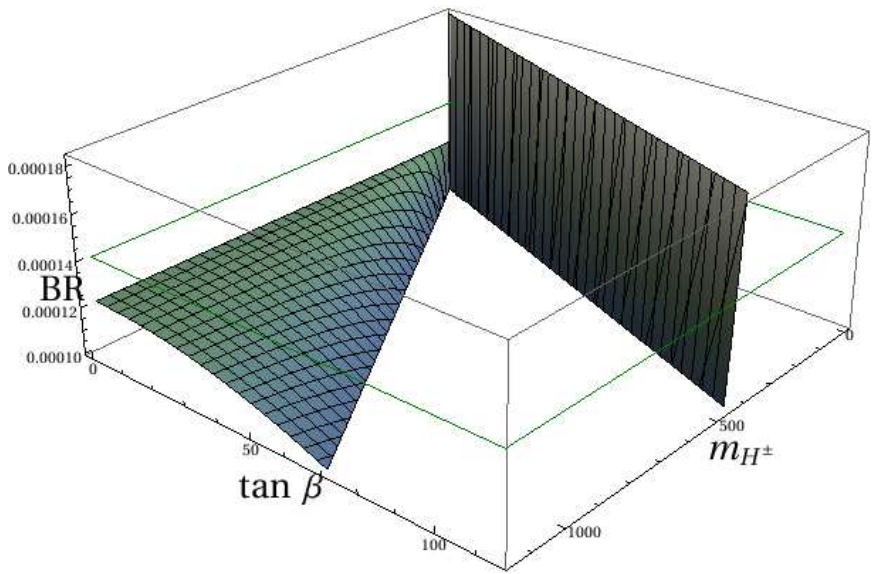,scale=0.6}}
\caption{$BR(B_u\to \tau^+\nu_{\tau})$ as a function of $\tan\beta$
and $m_{H^\pm}$. The plotted range of the BR corresponds to the 1-$\sigma$
range of the world average measurement $(1.42\pm 0.43)\cdot 10^{-4}$, and 
the line indicates the central value.}
\label{3Dplot}
\end{figure}

The effect of $H^\pm$ in the 2HDM (Model~II)
on the decays $B^\pm_u\to \ell^+\nu_\ell$ was considered in \cite{Hou:1992sy}
and the analogous analysis for $B^\pm_c\to \ell^+\nu_\ell$ was
presented in \cite{Du:1997pm}. In both cases the $H^\pm$ 
contribution modifies the SM prediction by a global factor $r^q_H$ where:
\begin{equation} \label{r_H}
r^q_H=[1-\tan^2\!\beta\,(M_{B_q}/m_{H^\pm})^2]^2\equiv [1-R^2M_{B_q}^2]^2
\end{equation}
The $H^\pm$ contribution interferes destructively with 
that of $W^\pm$. There are two solutions for 
$r^q_H=1$ which occur at $R=0$ and $R\sim 0.27$ GeV$^{-1}$
for $B^\pm_u\to \ell^+\nu_\ell$ ($R=0$ and $R\sim 0.26$ GeV$^{-1}$ 
for $B^\pm_c\to \ell^+\nu_\ell$).
This is shown in Fig.~\ref{3Dplot} where
$BR(B_u\to \tau^+\nu_{\tau})$ is plotted as a function of $\tan\beta$
and $m_{H^\pm}$. For $\tan\beta/m_{H^\pm}\sim 0$, the BR remains at its SM
value (slightly higher than the thin line indicating the
central value of the experimental measurement), but this
SM value can also be achieved along a line through the
steep part of the surface where $r_H=(1-2)^2=1$.

If the $b$ quark couples to both Higgs doublets at tree-level
(which is referred to as the type-III 2HDM),
Eq.~(\ref{r_H}) is modified to \cite{Akeroyd:2003zr}:
\begin{equation}
r_H=\left(1-\frac{\tan^2\beta}{1+\tilde\epsilon_0\,\tan\beta}\,
\frac{m^2_B} {m^2_{H^\pm}}\right)^{\!2}
\label{mod_r_H}
\end{equation}
In the Minimal Supersymmetric SM (MSSM),
the parameter $\tilde\epsilon_0$ does not appear at tree-level but is generated at 
the 1-loop level  \cite{Banks:1987iu,Hall:1993gn}
(with the main contribution originating from gluino diagrams)
and may reach values of 0.01. 
The redefinition of both the $b$ quark 
Yukawa coupling and the CKM matrix element  
$V_{ub}$ are encoded in $\tilde\epsilon_0$ 
\cite{Buras:2002vd,D'Ambrosio:2002ex}. 
The impact of $\tilde\epsilon_0\ne 0$ 
on $r_H$ has been developed in \cite{Itoh:2004ye,Chen:2006nua,Isidori:2006pk,Nierste:2008qe}.
In particular, the value of $R$ where $r_H=1$ shifts
depending on the magnitude and sign of $\tilde\epsilon_0$.

\begin{figure}
\center{\epsfig{file=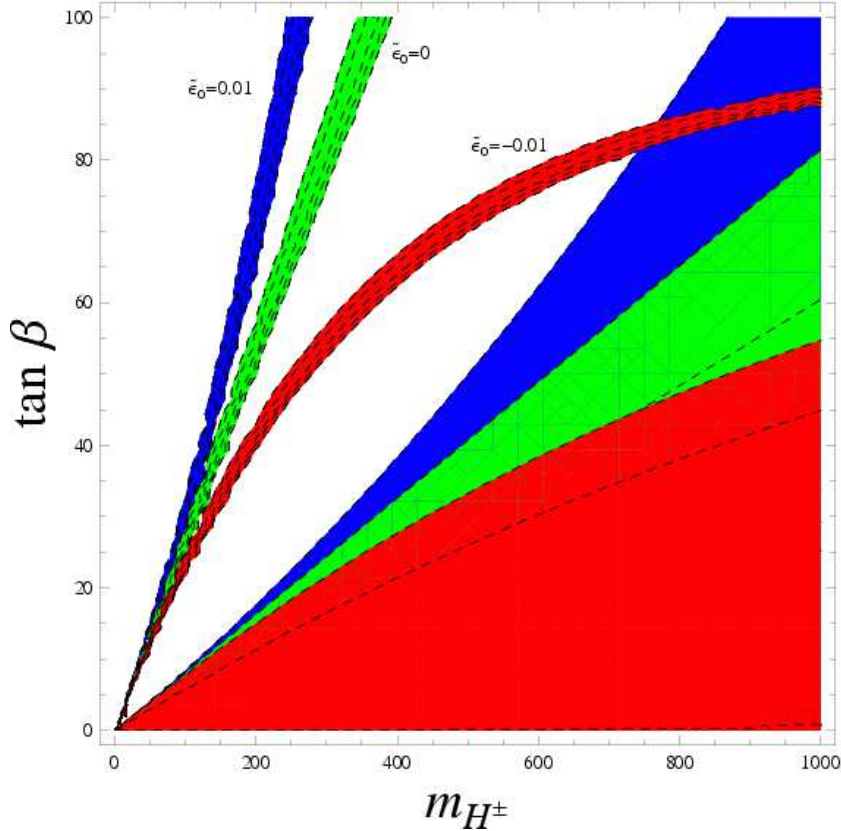,scale=0.65}}
\caption{The constraint on the $\tan\beta$-$m_{H^\pm}$ plane 
in the 2HDM (Type~III) from 
the measurement of $BR(B^\pm\to \tau^\pm\nu_{\tau})$. The coloured areas correspond
to allowed ranges of $\tan\beta$ and $m_{H^\pm}$ for various values of
$\tilde\epsilon_0=0,0.01,-0.01$ (green, blue and red, respectively),
$BR(B_u\to \tau^+\nu_{\tau})$ (1-$\sigma$ range, overlapping)
and $f_B$ (overlapping).}
\label{tanbeta-mH-plane}
\end{figure}

In Fig.~\ref{tanbeta-mH-plane} 
we show the impact of the measurement of $BR(B_u\to \tau^+\nu_{\tau})$
on the plane of $[\tan\beta$,$m_{H^\pm}]$ in the 2HDM (Type~III) which
updates the study of \cite{Akeroyd:2003zr}
(for a recent analogous plot with a somewhat lower value of $f_B$ see 
\cite{Kim:2007uq}). The white regions are excluded and 
the shaded areas correspond to $BR(B_u\to \tau^+\nu_{\tau})$
within the 1-$\sigma$ experimental range. We plot overlapping
bands for the  1-$\sigma$ ranges of the input parameters and consider
$\tilde\epsilon_0=0,0.01,-0.01$. In the MSSM, positive values for
$\tilde\epsilon_0$, corresponding to positive values of the
$\mu$-parameter, are preferred in order to explain the $(g-2)_\mu$ anomaly
\cite{Isidori:2007jw},
but in general both signs are possible. 

The different values for $\tilde\epsilon_0$ result 
in significantly different allowed regions
in the plane of $[\tan\beta$,$m_{H^\pm}]$. 
Importantly, these constraints from $BR(B_u\to \tau^+\nu_{\tau})$
are from a tree-level process and when applied to the MSSM 
are only sensitive to the assumptions for the soft SUSY breaking sector via $\tilde\epsilon_0$
(recently emphasized in \cite{Barenboim:2007sk}), 
i.e., a higher order effect.
In contrast, other important $B$ physics observables such as 
$b\to s\gamma$, $B_s-\overline B_s$ mixing and $B_{d,s}\to \mu\mu$
are all loop induced processes. Consequently, constraints on the
plane $[\tan\beta$,$m_{H^\pm}]$ from such processes are
very sensitive to the assumptions made for the sparticle masses,
and in certain cases the constraints can be removed completely
\footnote{In the non-SUSY 2HDM (Model~II) $b\to s\gamma$
constrains $m_{H^\pm}$ independently of $\tan\beta$. A recent
study \cite{Misiak:2006zs,Misiak:2006ab} obtains $m_{H^\pm}>295$ GeV.}. 
In global studies of $B$ physics observables
in specific MSSM scenarios \cite{Isidori:2006pk,Carena:2007aq}
the measured $BR(B_u\to \tau^+\nu_{\tau})$
also plays an important role. Certainly, improved precision for
$BR(B_u\to \tau^+\nu_{\tau})$ is desirable and very relevant in the
era of the LHC in which the plane $[\tan\beta$,$m_{H^\pm}]$
will be probed via direct production of Higgs bosons.
Currently only High Luminosity $B$ factories operating at the
$\Upsilon(4S)$ are discussed when considering future facilities which could
offer improved precision for $BR(B_u\to \tau^+\nu_{\tau})$.

Another promising approach to probe the plane $[\tan\beta$,$m_{H^\pm}]$
is via the tree-level $H^\pm$ contribution to the semileptonic decays
$B\to D\tau \nu$  
\cite{Itoh:2004ye, Chen:2006nua, Nierste:2008qe,Grzadkowski:1992qj}.
We note here that $H^\pm$ can mediate the analogous leptonic decays 
$K^\pm\to \mu^\pm\nu$ \cite{Hou:1992sy,Antonelli:2008jg} and
$D^\pm_s\to \mu^\pm\nu,\tau^\pm\nu$ \cite{Hou:1992sy,Hewett:1995aw} 
but constraints on the plane $[\tan\beta$,$m_{H^\pm}]$ from 
these processes are not yet competitive.
However, such processes might play a role in 
the future with increased experimental precision and
reduced theoretical uncertainties.

\boldmath
\section{At the $Z$ peak}
\unboldmath

In this section we discuss the searches for $B^\pm\to \tau^\pm\nu$
using data from $e^+e^-$ collisions at the $Z$ peak ($\sqrt s \sim 91$ GeV).
It was pointed out in \cite{Mangano:1997md} that such searches
would also be sensitive to the decay $B^\pm_c\to \tau^\pm\nu$.
Assuming that the detection efficiencies are the 
same\footnote{In practice, the shorter lifetime of $B^\pm_c$ would
result in a slightly inferior detection efficiency \cite{Mangano:1997md}.} 
the ratio of $\tau^\pm\nu$ events originating from $B^\pm\to \tau^\pm\nu$
and $B^\pm_c\to \tau^\pm\nu$ is given by:
\beq
\frac{N_c}{N_u}=\left|\frac{V_{cb}}{V_{ub}}\right|^2\frac{f(b\to B^\pm_c)}
{f(b\to B^\pm)} \left(\frac{f_{B_c}}{f_B}\right)^{\!2}
\frac{M_{B_c}}{M_B}\frac{\tau_{B_c}}{\tau_B}
\frac{\left(1-\frac{m^2_\tau} {M^2_{B_c}}\right)^{\!2}}
     {\left(1-\frac{m^2_\tau} {M^2_{B}}\right)^{\!2}}
\label{Nc}
\eeq
The largest uncertainty in the determination of $N_c$ is from
the transition
probability  $f(b\to B^\pm_c)$ and the decay constant $f_{B_c}$.
The magnitude of $N_c$ is suppressed by the small 
 $f(b\to B^\pm_c)$ but this can be compensated by
the large ratio $(V_{cb}f_{B_c})^2/(V_{ub}f_{B})^2$. Consequently
$N_c$ can be similar in magnitude to $N_u$.
In the analysis of \cite{Mangano:1997md} three scenarios were defined in order
to account for the error in the determination of $N_c/N_u$:
``Central'' and ``Max/Min'' ($\pm 1\sigma$ above/below the central values
of the input parameters).
Since the analysis of \cite{Mangano:1997md} there have been significant
improvements in the measurements of $V_{ub}$ and $V_{cb}$.
In addition, the decay constant $f_B$ has now been calculated in 
unquenched lattice QCD
with smaller errors and a central value considerably larger
\cite{Gray:2005ad}
than the values used  in both \cite{Mangano:1997md} and 
the L3 analysis \cite{Acciarri:1996bv}.
We are unaware of an unquenched lattice QCD calculation of $f_{B_c}$ 
and the error in this parameter has not been reduced significantly since 
\cite{Mangano:1997md}.
The main uncertainty in the ratio $N_c/N_u$ is from 
$f(b\to B^\pm_c$), which in \cite{Mangano:1997md} 
was varied in the range suggested by theoretical estimations
\cite{Lusignoli:1991bn}:
$2\times 10^{-4}<f(b\to B^\pm_c)< 1\times 10^{-3}$. 
At that time $B^\pm_c$ was still undiscovered and hence there was
no measurement of $f(b\to B^\pm_c$).

\begin{figure}[htbp]
\includegraphics*[width=3.5 in]{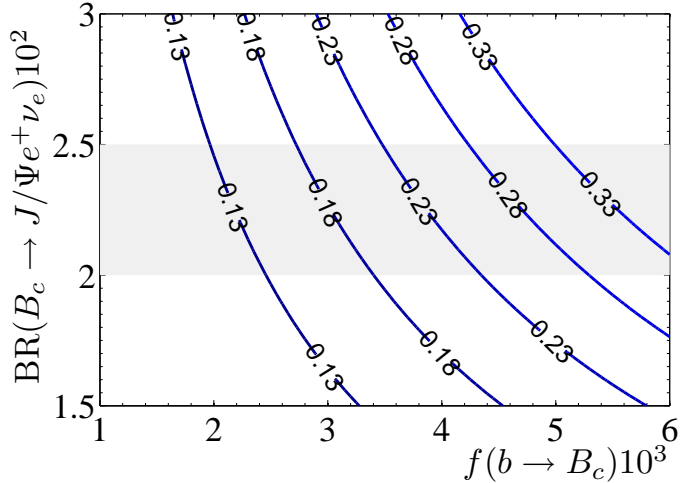}
\caption{Contours of ${\cal R}_{e}$ in the plane of BR($B_c\to
J/\Psi e^{+} \nu_{e}$) and transition probability $f(b\to B_c)$. The
shaded region denotes the theoretical prediction for BR($B_c\to
J/\Psi e^{+} \nu_{e}$).}
 \label{fig:re}
\end{figure}

However, $f(b\to B^\pm_c)$ can now be extracted 
(although with a large uncertainty) from the measurement of ratio of $B_c^\pm\to
J/\Psi \ell^{+} \nu_{\ell}$ to  $B^\pm\to J/\Psi K^{\pm}$ which is
defined by:
 \beq
{\cal R}_{\ell} = \frac{\sigma(B_c^+)\cdot {\rm BR}(B_c\to J/\psi
\ell^\pm \nu_{\ell})}
   {\sigma(B^+)\cdot {\rm BR}(B\to J/\psi K^+)}
\label{tevatron98}
 \eeq
Tevatron Run II data gives ${\cal R}_{e}=0.28\pm 0.07$
\cite{Abulencia:2006zu}, and the denominator in eq.~(\ref{tevatron98})
has been measured precisely by various experiments.
The transition probability
$f(b\to B_c)$ determines $\sigma(B_c^+)$ and several 
theoretical calculations are available for BR($B_c\to J/\psi
\ell^\pm \nu_{\ell})$). In Fig.~\ref{fig:re}
we display contours of ${\cal
R}_{e}$ as a function of ${\rm BR}(B_c\to J/\Psi e^{+} \nu_{e})$ and
$f(b\to B_c)$, and the band denotes the
prediction of the various theoretical calculations
for ${\rm BR}(B_c\to J/\Psi e^{+}
\nu_{e})$ whose values lie in the range $(2.0\sim 2.5)\%$ \cite{theory_Bc}.
From Fig.~\ref{fig:re} one can see that the Tevatron Run I measurement
of ${\cal R}_{e}=0.13\pm 0.05$ \cite{Abe:1998fb} is accommodated
by $f(b\to B_c)=1.3\times 10^{-3}$. However, in order to
satisfy the central value of the Run II measurement,
the transition probability $f(b\to B_c$) needs to be $4.5 \times 10^{-3}$.
An even larger value for $f({b\to B_c})$ was suggested in
Ref.~\cite{Gershtein:2007sm}. Such unexpectedly large values of $f(b\to B_c)$,
which are indicated by Tevatron Run II data, would significantly
enhance the contribution of $B_c^\pm\to \tau^\pm\nu$ to the
LEP searches for $B^\pm\to \tau^\pm\nu$.
Of course, $f(b\to B_c)$ is dependent on the available centre of
mass energy (at higher energies there is more phase space to
produce a charm quark instead of a light quark), but the value
of $f(b\to B_c)$ is expected to be of comparable size at LEP and at the
LHC \cite{Lusignoli:1991bn}.
In our numerical analysis in Sec.~\ref{lepsearch},
we will consider values of $f(b\to B^\pm_c)$ 
up to $5\cdot 10^{-3}$.

\boldmath
\subsection{The LEP search for $B^\pm\to \tau^\pm\nu$ and
the contribution of $B^\pm_c\to \tau^\pm\nu$ \label{lepsearch}}
\unboldmath
Three LEP collaborations searched for the decay $B^\pm\to \tau^\pm\nu$
using data taken at the $Z$ peak ($\sqrt s=91$ GeV).
L3 \cite{Acciarri:1996bv} used around $1.5\times 10^6$ hadronic decays of
the $Z$ boson which 
corresponds to about half their total data \cite{:2005ema}.
DELPHI \cite{Abreu:1999xe} and ALEPH 
\cite{Barate:2000rc} used their full data samples of
around $3.6\times  10^6$ hadronic decays of the $Z$ boson. The best sensitivity
was from the L3 experiment which set the upper limit 
BR($B^\pm\to \tau^\pm\nu)<5.7\times 10^{-4}$. 
The L3 limit is of particular interest since it could be improved if the
full data sample of $\sim 3.6\times 10^6$ hadronic $Z$ boson decays were used. 

The LEP searches were sensitive to $\tau^\pm\nu$ events originating
from both $B^\pm\to \tau^\pm\nu$ and $B^\pm_c\to \tau^\pm\nu$.
Hence the published limits constrain the  
``effective branching ratio'' defined by:
\beq
{\rm BR}_{\rm eff}={\rm BR}(B^\pm\to \tau^\pm \nu)\left(1+\frac{N_c}
 {N_u}\right)   \label{BReff}
\eeq
This expression applies to searches for $B^\pm\to \tau^\pm \nu$ at the
$Z$ peak. For searches at the $\Upsilon(4S)$ clearly $N_c=0$ 
and ${\rm BR}_{\rm eff}$=BR($B^\pm\to \tau^\pm\nu$).
In our numerical
analysis in this section we will use eq.~(\ref{BReff})
with the experimental value of  BR($B^\pm\to \tau^\pm\nu$)
as input. The calculation of $N_c/N_u$ in eq.~(\ref{BReff})
uses eq~.(\ref{Nc}) (i.e. the expression for the SM) 
with input parameters taken from Table~\ref{parameters}.
Our analysis can be applied to any model
for which $N_c/N_u \sim |N_c/N_u|_{\rm SM}$, which
includes the 2HDM because the scale factors in eq.~(\ref{r_H})
are almost equal.

The ``max'' scenario of \cite{Mangano:1997md} showed that the current
limit of ${\rm BR}_{\rm eff}<5.7\times 10^{-4}$ would be sensitive
to the SM rate for BR$(B^\pm\to \tau^\pm \nu)$.
The measurements of 
BR($B^\pm\to \tau^\pm\nu)$ at the $B$ factories are consistent with the 
SM prediction which suggests that the L3 search was not so far from 
observing a signal.

\begin{figure}
\center{\epsfig{file=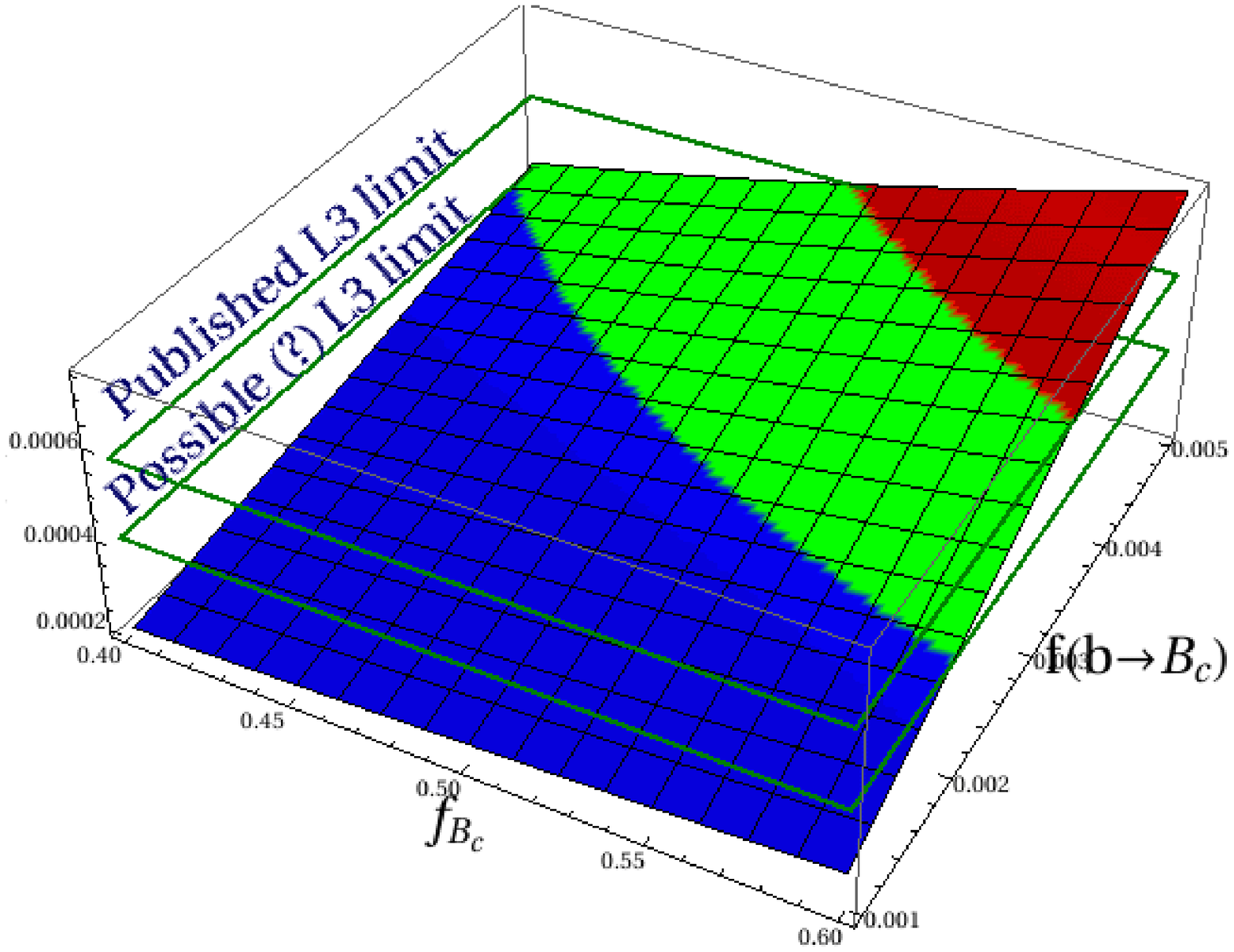,scale=0.7}
\epsfig{file=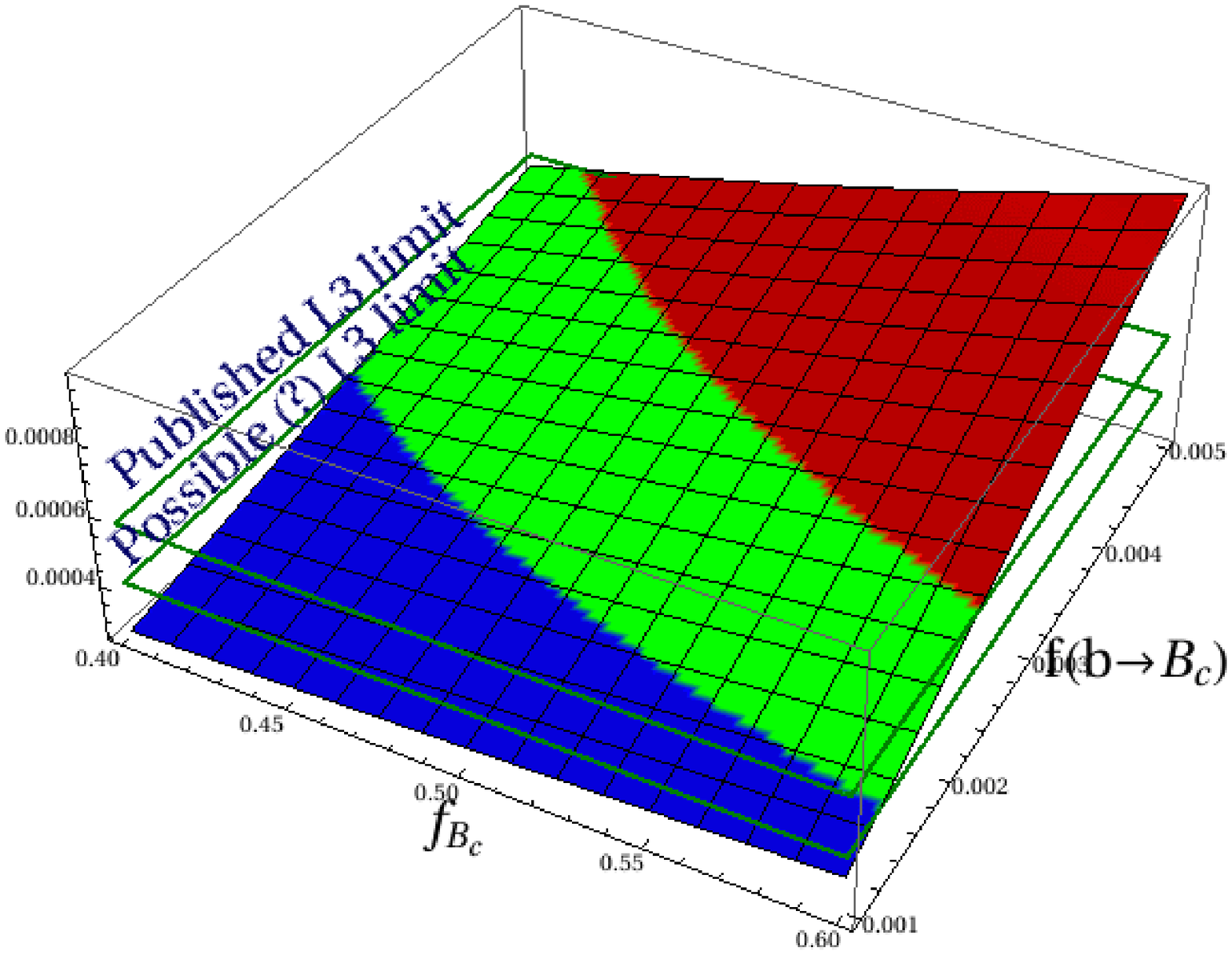,scale=0.7}}
\caption{The {\em effective} $BR(B^\pm\to \tau^\pm\nu)$ at the $Z$
peak in the plane $[f(b\to B_c^\pm), f_{B_c}]$.
The published L3 limit and a possible stricter limit are indicated.
In the upper (lower) panel ${\rm BR}(B^\pm\to \tau^\pm \nu)$ is taken to be 
the central value of the world average measurement 
(the 1-$\sigma$ upper value).
\label{BReff-fig}}
\end{figure}
In Fig.~\ref{BReff-fig} we plot ${\rm BR}_{\rm eff}$ in the plane 
$[f(b\to B_c^\pm), f_{B_c}]$, 
for two values of BR$(B^\pm\to \tau^\pm \nu)$ corresponding to the
central value and $1\sigma$ above the world average.
All other parameters are held at their central values from
Tab.~\ref{parameters}.
The region above the contour ${\rm BR_{eff}}=5.7\times 10^{-4}$ (red/dark grey)
is excluded by the L3 limit \cite{Acciarri:1996bv}, while the 
contour ${\rm BR_{\rm eff}}=4\times 10^{-4}$ represents the 
hypothetical sensitivity  if the full data of $3.6\times  
10^6$ hadronic decays of the $Z$ boson were used.
The green/light grey area between the two contours is the area
where a signal would be seen if the full dataset were studied.
Depending on the other input parameters and the $B^\pm\to \tau^\pm\nu$
branching ratio, this area can cover a very significant part of
the $[f(b\to B_c^\pm), f_{B_c}]$
parameter space. We therefore consider a re-analysis using
the full L3 dataset very worthwhile.

A different way of studying the number of $B_c$ events 
was followed in \cite{Mangano:1997md}. The number of $B_c$
events per $B_u$ event can be calculated as a function of
$f(b\to B^\pm_c)$, and the authors obtained $N_c/N_u=1.2 f(b\to B^\pm_c)/10^{-3}$
for central values of the input parameters (``max'' scenario: 2.3).
With updated values for the input parameters, we now find
\beq
\frac{N_c}{N_u}=\quad \parbox{7cm}
  { \begin{tabular}{ccl}
    $0.48$ & $\cdot {f_{b\to B_c}/10^{-3}} $ & (central values) \\
    $1.50$ & $\cdot {f_{b\to B_c}/10^{-3}} $ & (``optimistic'' values)
    \end{tabular} }
\eeq
where for the ``optimistic'' values of the parameters from
Tab.~\ref{parameters} we have chosen that end of the 1-$\sigma$
range that results in a higher value for $N_c/N_u$, and the ``optimistic''
$f_{B_c}$ was chosen to be 550 MeV.

These numbers are lower than those of \cite{Mangano:1997md}
mainly because the central value of 
$V_{ub}/V_{cb}$ has increased in the last ten years. 
The inverse of this ratio enters $N_c/N_u$ quadratically
and therefore reduces this quantity. On the other hand,
experimental data does not preclude values of $f(b\to B^\pm_c)$
which are much higher (a few $\times 10^{-3}$) 
than the theoretical estimates, 
and so the admixture of $B_c^\pm\to \tau^\pm \nu$ can still easily reach 100\%.

\boldmath
\subsection{Giga $Z$ option at a future $e^+e^-$ Linear Collider}
\unboldmath

A future $e^+e^-$ Linear Collider operating at the $Z$ peak
with a luminosity of $5\times 10^{33}cm^{-2}s^{-1}$
could produce $10^9$ $Z$ bosons in $50-100$ days of operation
\cite{Hawkings:1999ac,
AguilarSaavedra:2001rg,Ali:2000cq,Djouadi:2007ik}) 
This corresponds to roughly 1000 times the number of $Z$ bosons recorded 
at each LEP detector. Historically, limits on 
$B^\pm\to \tau^\pm\nu$ from $Z$ decays have been comparable to
(if not stronger than) those at $\Upsilon(4S)$ for the same number 
of $Z$ bosons and $B$ mesons. For example, the 
CLEO collaboration obtained BR($B^\pm\to \tau^\pm\nu)< 8.4\times 10^{-4}$
with $9.7\times 10^6$ $B$ mesons \cite{Browder:2000qr}, while
L3 obtained BR($B^\pm\to \tau^\pm\nu)< 5.7\times 10^{-4}$
with $1.5\times 10^6$ hadronic decays of the $Z$ boson.

High luminosity $B$ factories 
\cite{Yamauchi:2002ru,Bigi:2004kn,Hashimoto:2004sm,Browder:2004wu,
Hewett:2004tv,Bona:2007qt,Browder:2008em} 
anticipate data samples of $10^{10}$ $B$
mesons. 
By the time of operation of a Giga $Z$ the two main sources of
uncertainty in $N_c$ (and hence ${\rm BR_{eff}}$) will have been substantially reduced.
The error in $f(b\to B_c$) will be reduced from LHC-b measurements
\cite{Ball:2000ba} of the cross-section in Eq.~\ref{tevatron98}, and 
improved lattice calculations
of $f_{B_c}$ and/or ($f_{B_c}/f_B$) would also reduce the error in $N_c$.
In Table \ref{GigaZ} we present the required number of $B$ mesons and $Z$ bosons
for a precision of $20\%$ and $4\%$ in the measurement of $B^\pm\to \tau^\pm\nu$
at a high luminosity $B$ factory and ${\rm BR_{eff}}$ at Giga $Z$. 
The numbers for a high luminosity $B$ factory are taken from \cite{Bona:2007qt}.
For the Giga $Z$ precision we assume a signal of ${\rm BR_{eff}}=4\pm 2\times 10^{-4}$ ($50\%$ error) at L3 with 
$3.6\times 10^6$ hadronic $Z$ decays, and scale the error by $1/\sqrt N$, where $N$ is the total
number of $Z$ bosons at Giga $Z$ divided by the full L3 data sample of $\sim 5.1\times 10^6$ $Z$ bosons.

\begin{table}
\center{\begin{tabular}{|c|c|c|}
\hline
Error BR($B^\pm/B^\pm_c\to \tau^\pm\nu$) &  High Lum. $B$ Factory ($B$ mesons)  & Giga $Z$ ($Z$ bosons) \\\hline
$20\%$  & $2.2\times 10^{9}$ & $3.2\times 10^7$       \\
$4\%$ & $8.1\times 10^{10}$ & $8\times 10^{8}$ \\
\hline
\end{tabular} }
\caption{Required number of $B$ mesons ($Z$ bosons) for a precision of
$20\%$ and $4\%$ in the measurement of 
BR($B^\pm/B^\pm_c\to \tau^\pm\nu$), assuming a signal of ${\rm BR_{eff}}=4\pm 2\times 10^{-4}$
at L3.}
\label{GigaZ}
\end{table}

It is clear from Table~\ref{GigaZ} that
a Giga $Z$ facility might be capable of measuring ${\rm BR}_{\rm eff}$
in Eq.~\ref{BReff} with similar precision to that anticipated for
$B^\pm\to \tau^\pm\nu$ at high luminosity $B$ factories. 
We believe that this competitiveness of the Giga $Z$ facility has 
not been pointed out for the leptonic $B$ decays 
although it has been emphasized for the
decay $B \to X_s \nu \overline\nu$ in \cite{Ali:2000cq}.
If both facilities
were realized this would enable
competitive and complementary constraints on $\tan\beta/m_{H^\pm}$ 
in the context of models with $H^\pm$. 

\section{Conclusions}

The decay $B^\pm\to \tau^\pm\nu$ has been observed 
at the $e^+e^-$ $B$ factories and is recognized as an
important constraint on the parameter $\tan\beta/m_{H^\pm}$ in the
context of models with Two Higgs doublets.
We studied the contribution of $B_c^\pm\to \tau^\pm\nu$ to the
LEP searches for $B^\pm\to \tau^\pm\nu$ 
(first pointed out in \cite{Mangano:1997md}), whose main uncertainty 
is from the value for the transition probability $b\to B^\pm_c$ which is
now being measured at the Tevatron Run II. 
Using values of this transition probability which are 
consistent with the current Tevatron measurements
(which accommodate values significantly larger than the 
theoretical estimations), 
we found that the contribution 
of $B_c^\pm$ mesons to the search for $B^\pm/B_c^\pm\to \tau^\pm\nu$
can be as large as that of $B^\pm$. 
We suggested that a re-analysis of the L3 search 
for $B^\pm\to \tau^\pm\nu$ \cite{Acciarri:1996bv}
using all the data taken at the $Z$ peak could provide 
a signal for the admixture of $B^\pm/B^\pm_c\to \tau^\pm\nu$.
Finally, it was pointed out that the Giga $Z$ option of
a future $e^+e^-$ collider could offer
measurements of these leptonic
$B^\pm/B^\pm_c$ decays which are comparable in precision 
and complementary with those anticipated at
the proposed high luminosity $B$ factories.

\section*{Acknowledgements}
A.G.A. was supported by National Cheng Kung University Grant 
No.\ OUA 95-3-2-057.


\begin{thebibliography}{99}

\bibitem{Ikado:2006un}
  K.~Ikado {\it et al.},
  Phys.\ Rev.\ Lett.\  {\bf 97}, 251802 (2006).

\bibitem{:2007bx}
  B.~Aubert {\it et al.}  [BABAR Collaboration],
Phys.\ Rev.\  D {\bf 76}, 052002 (2007).


\bibitem{:2007xj}
  B.~Aubert {\it et al.}  [BABAR Collaboration],
  Phys.\ Rev.\  D {\bf 77}, 011107 (2008).


\bibitem{HFAG}
Heavy Flavor Averaging Group, http://www.slac.stanford.edu/xorg/hfag/


\bibitem{Yamauchi:2002ru}
  M.~Yamauchi,
  Nucl.\ Phys.\ Proc.\ Suppl.\  {\bf 111}, 96 (2002).

\bibitem{Bigi:2004kn}
  I.~I.~Bigi and A.~I.~Sanda,
  arXiv:hep-ph/0401003.

\bibitem{Hashimoto:2004sm}
  S.~Hashimoto {\it et al.},``Letter of intent for KEK Super B Factory,'',
 KEK-REPORT-2004-4;
  A.~G.~Akeroyd {\it et al.}  [SuperKEKB Physics Working Group],
  arXiv:hep-ex/0406071.

\bibitem{Browder:2004wu}
  T.~E.~Browder and A.~Soni,
  Pramana {\bf 63}, 1171 (2004).


\bibitem{Hewett:2004tv}
  J.~L.~.~Hewett {\it et al.},
  arXiv:hep-ph/0503261.

\bibitem{Bona:2007qt}
  M.~Bona {\it et al.},
  arXiv:0709.0451 [hep-ex].

\bibitem{Browder:2008em}
  T.~E.~Browder, T.~Gershon, D.~Pirjol, A.~Soni and J.~Zupan,
  arXiv:0802.3201 [hep-ph].


\bibitem{Albright:1979yc}
  C.~H.~Albright, J.~Smith and S.~H.~H.~Tye,
  Phys.\ Rev.\  D {\bf 21}, 711 (1980).

\bibitem{Hou:1992sy}
W.~S.~Hou,
Phys.\ Rev.\ D {\bf 48}, 2342 (1993).


\bibitem{Acciarri:1996bv}
M.~Acciarri {\it et al.}  [L3 Collaboration],
Phys.\ Lett.\ B {\bf 396}, 327 (1997).

\bibitem{Abreu:1999xe}
P.~Abreu {\it et al.}  [DELPHI Collaboration],
Phys.\ Lett.\ B {\bf 496}, 43 (2000).

\bibitem{Barate:2000rc}
R.~Barate {\it et al.}  [ALEPH Collaboration],
Eur.\ Phys.\ J.\ C {\bf 19}, 213 (2001).

\bibitem{Mangano:1997md}
M.~L.~Mangano and S.~R.~Slabospitsky,
Phys.\ Lett.\ B {\bf 410}, 299 (1997).


\bibitem{Abe:1998fb}
F.~Abe {\it et al.}  [CDF Collaboration],
Phys.\ Rev.\ D {\bf 58}, 112004 (1998).

\bibitem{Abulencia:2005usa}
  A.~Abulencia {\it et al.}  [CDF Collaboration],
  Phys.\ Rev.\ Lett.\  {\bf 96}, 082002 (2006)

\bibitem{Abulencia:2006zu}
  A.~Abulencia {\it et al.}  [CDF Collaboration],
  Phys.\ Rev.\ Lett.\  {\bf 97}, 012002 (2006)

\bibitem{Aaltonen:2007gv}
  T.~Aaltonen {\it et al.}  [CDF Collaboration],
  arXiv:0712.1506 [hep-ex].

\bibitem{:2005ema}
  ``Precision electroweak measurements on the Z resonance,''
  Phys.\ Rept.\  {\bf 427}, 257 (2006).


\bibitem{Hawkings:1999ac}
  R.~Hawkings and K.~Monig,
  Eur.\ Phys.\ J.\ direct C {\bf 1}, 8 (1999).


\bibitem{Ali:2000cq}
  A.~Ali, D.~Benson, I.~I.~Y.~Bigi, R.~Hawkings and T.~Mannel,
  arXiv:hep-ph/0012218.

\bibitem{AguilarSaavedra:2001rg}
  J.~A.~Aguilar-Saavedra {\it et al.}  [ECFA/DESY LC Physics Working Group],
  arXiv:hep-ph/0106315.

\bibitem{Djouadi:2007ik}
  A.~Djouadi, J.~Lykken, K.~Monig, Y.~Okada, M.~J.~Oreglia and S.~Yamashita,
  ``International Linear Collider Reference Design Report Volume 2: PHYSICS AT
  THE ILC,''
  arXiv:0709.1893 [hep-ph].


\bibitem{Du:1997pm}
D.~S.~Du, H.~Y.~Jin and Y.~D.~Yang,
Phys.\ Lett.\ B {\bf 414}, 130 (1997).


\bibitem{Akeroyd:2003zr}
  A.~G.~Akeroyd and S.~Recksiegel,
  J.\ Phys.\ G {\bf 29}, 2311 (2003)

\bibitem{Banks:1987iu}
  T.~Banks,
  Nucl.\ Phys.\  B {\bf 303}, 172 (1988).

\bibitem{Hall:1993gn}
  L.~J.~Hall, R.~Rattazzi and U.~Sarid,
  Phys.\ Rev.\  D {\bf 50}, 7048 (1994).


\bibitem{Buras:2002vd}
    A.~J.~Buras, P.~H.~Chankowski, J.~Rosiek and L.~Slawianowska,
  Phys.\ Lett.\  B {\bf 546}, 96 (2002);
  Nucl.\ Phys.\  B {\bf 659}, 3 (2003).

\bibitem{D'Ambrosio:2002ex}
  G.~D'Ambrosio, G.~F.~Giudice, G.~Isidori and A.~Strumia,
  Nucl.\ Phys.\  B {\bf 645}, 155 (2002).


\bibitem{Itoh:2004ye}
  H.~Itoh, S.~Komine and Y.~Okada,
  Prog.\ Theor.\ Phys.\  {\bf 114}, 179 (2005).

\bibitem{Isidori:2006pk}
  G.~Isidori and P.~Paradisi,
  Phys.\ Lett.\  B {\bf 639}, 499 (2006).

\bibitem{Chen:2006nua}
  C.~H.~Chen and C.~Q.~Geng,
  JHEP {\bf 0610}, 053 (2006)

\bibitem{Nierste:2008qe}
  U.~Nierste, S.~Trine and S.~Westhoff,
  arXiv:0801.4938 [hep-ph].

\bibitem{Kim:2007uq}
  C.~S.~Kim and R.~M.~Wang,
  arXiv:0712.2954 [hep-ph].

\bibitem{Isidori:2007jw}
  G.~Isidori, F.~Mescia, P.~Paradisi and D.~Temes,
  Phys.\ Rev.\  D {\bf 75}, 115019 (2007)


\bibitem{Barenboim:2007sk}
  G.~Barenboim, P.~Paradisi, O.~Vives, E.~Lunghi and W.~Porod,
  arXiv:0712.3559 [hep-ph].

\bibitem{Misiak:2006zs}
  M.~Misiak {\it et al.},
  Phys.\ Rev.\ Lett.\  {\bf 98}, 022002 (2007).

\bibitem{Misiak:2006ab}
  M.~Misiak and M.~Steinhauser,
  Nucl.\ Phys.\  B {\bf 764}, 62 (2007).

\bibitem{Carena:2007aq}
  M.~S.~Carena, A.~Menon and C.~E.~M.~Wagner,
  Phys.\ Rev.\  D {\bf 76}, 035004 (2007);
  J.~R.~Ellis, S.~Heinemeyer, K.~A.~Olive, A.~M.~Weber and G.~Weiglein,
  JHEP {\bf 0708}, 083 (2007);
  M.~Albrecht, W.~Altmannshofer, A.~J.~Buras, D.~Guadagnoli and D.~M.~Straub,
  JHEP {\bf 0710}, 055 (2007);
  J.~Ellis, J.~S.~Lee and A.~Pilaftsis,
  Phys.\ Rev.\  D {\bf 76}, 115011 (2007);
  F.~Domingo and U.~Ellwanger,
  JHEP {\bf 0712}, 090 (2007).

\bibitem{Grzadkowski:1992qj}
  B.~Grzadkowski and W.~S.~Hou,
  Phys.\ Lett.\  B {\bf 283}, 427 (1992);
  M.~Tanaka,
  Z.\ Phys.\  C {\bf 67}, 321 (1995);
  K.~Kiers and A.~Soni,
  Phys.\ Rev.\  D {\bf 56}, 5786 (1997);
  J.~F.~Kamenik and F.~Mescia,
  arXiv:0802.3790 [hep-ph].


\bibitem{Antonelli:2008jg}
  M.~Antonelli {\it et al.}  [FlaviaNet Workshop on Kaon Decays
                  Collaboration],
  arXiv:0801.1817 [hep-ph];
  G.~Isidori,
  arXiv:0801.3039 [hep-ph].



\bibitem{Hewett:1995aw}
  J.~L.~Hewett,
  arXiv:hep-ph/9505246;
  A.~G.~Akeroyd,
  Prog.\ Theor.\ Phys.\  {\bf 111}, 295 (2004);
  A.~G.~Akeroyd and C.~H.~Chen,
  Phys.\ Rev.\  D {\bf 75}, 075004 (2007);
  J.~L.~Rosner and S.~Stone,
  arXiv:0802.1043 [hep-ex].

\bibitem{Gray:2005ad}
  A.~Gray {\it et al.}  [HPQCD Collaboration],
  Phys.\ Rev.\ Lett.\  {\bf 95}, 212001 (2005).

\bibitem{Lusignoli:1991bn}
  M.~Lusignoli, M.~Masetti and S.~Petrarca,
  Phys.\ Lett.\  B {\bf 266}, 142 (1991).

\bibitem{theory_Bc}C.~H. Chang and Y.~Q. Chen, Phys. Rev. D{\bf 49}, 3399
(1994); A. Abd El-Hady, J.~H. Mu${\rm \tilde{n}}$oz and J.~P. Vary,
Phys. Rev. D{\bf 62}, 014019 (2000); V.~V. Kiselev, A.K. Likhoded
and A.~I. Onishchenko, Nucl. Phys. B{\bf 569}, 473 (2000); V.~V.
Kiselev, A.~E. Kovalsky and A.~K. Likhoded, Nucl. Phys. B{\bf 585},
353 (2000);  M.~A. Ivanov, J.~G. Korner and P. Santorelli, Phys.
Rev. D{\bf 63},  074010 (2001).

\bibitem{Gershtein:2007sm}
  S.~S.~Gershtein and A.~K.~Likhoded,
  arXiv:0706.0963 [hep-ph].

\bibitem{Browder:2000qr}
  T.~E.~Browder {\it et al.}  [CLEO Collaboration],
  Phys.\ Rev.\ Lett.\  {\bf 86}, 2950 (2001).

\bibitem{Ball:2000ba}
  P.~Ball {\it et al.},
  arXiv:hep-ph/0003238.





\end{thebibliography}
\end{document}